\documentclass[10pt, a4paper, twocolumn]{article} 
\usepackage{microtype} 

\usepackage{amsmath,amsfonts,amsthm} 

\usepackage[svgnames]{xcolor} 

\usepackage{booktabs} 

\usepackage{lastpage} 

\usepackage{graphicx} 

\usepackage{enumitem} 
\setlist{noitemsep} 

\usepackage{sectsty} 
\allsectionsfont{\usefont{OT1}{phv}{b}{n}} 

\usepackage{geometry} 

\usepackage[T1]{fontenc} 
\usepackage[utf8]{inputenc} 

\usepackage{XCharter} 

\usepackage{fancyhdr} 
\fancypagestyle{firstpage}{ 
  \rhead{\color{white} Right Header for just the first page}
}

\usepackage{titling} 

\newcommand{\HorRule}{\color{black}\rule{\linewidth}{1pt}} 

\pretitle{
	\vspace{-20em} 
	\includegraphics[width=1.5\textwidth 
	\hspace{-31pt} \vspace{-35pt}
	]{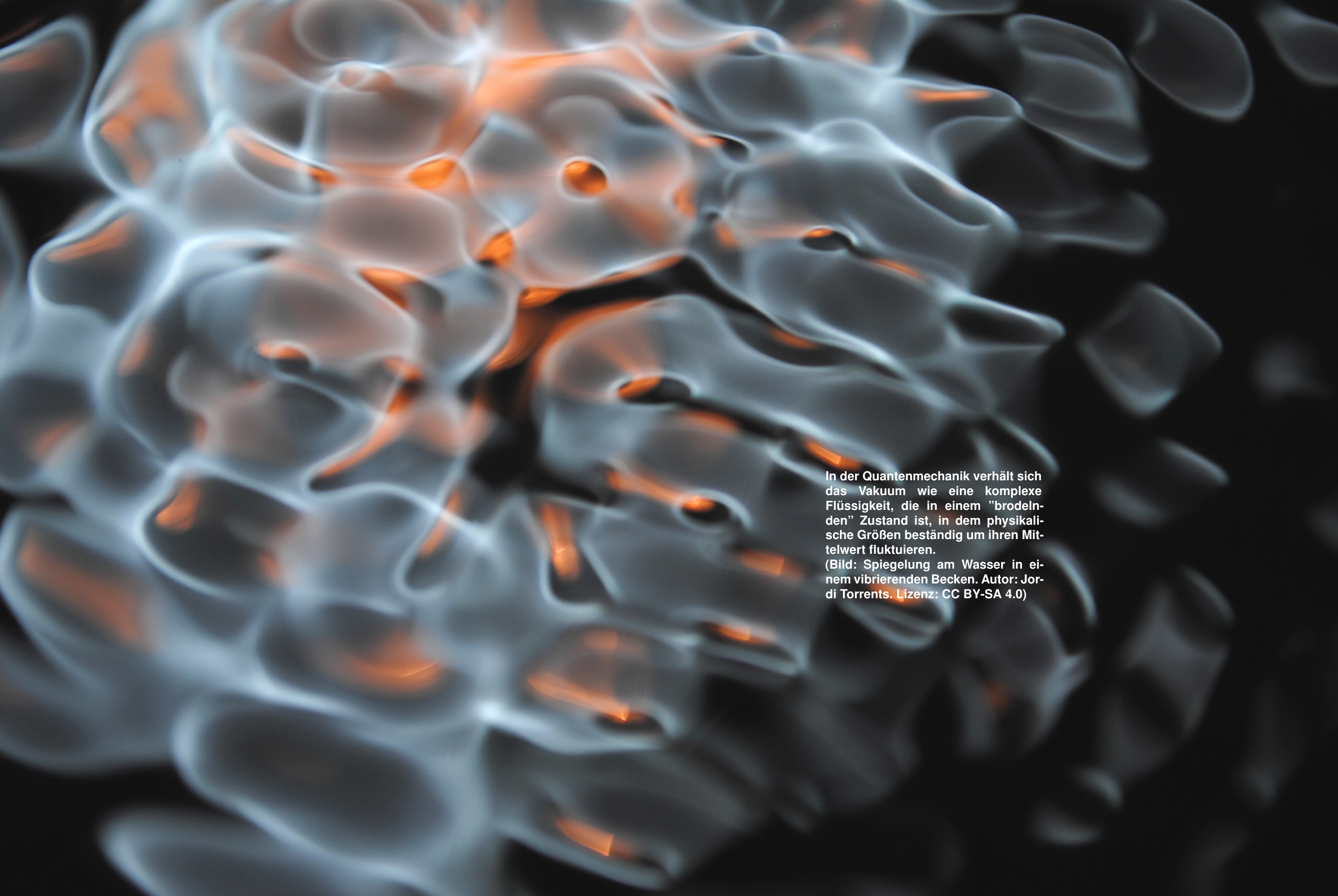}
	\fontsize{32}{36}\usefont{OT1}{phv}{b}{n}\selectfont 
	\color{white} 
}

\posttitle{\par\vskip 35pt  
} 

\preauthor{} 

\postauthor{ 
	\vspace{10pt} 
}

\usepackage{lettrine} 
\usepackage{fix-cm}	

\usepackage{xstring} 

\usepackage{amsmath,amsfonts,amsthm} 

\usepackage[format=plain, small, labelfont=bf, up, textfont={}, labelsep=space]{caption} 
\usepackage{graphicx} 

\usepackage[german]{babel} 

\title{
In der \emph{Unruhe} liegt die Kraft\\ 
\vspace{0.2em}\large{}Ein Blick auf Quantenfluktuationen und ihre Konsequenzen} 

\author{
	Francesco Intravaia$^{1}$, Daniel Reiche$^{1,2}$, Kurt Busch$^{1,2}$ 
	\\ 
	\footnotesize{$^{1}$Humboldt-Universität zu Berlin, Institut für Physik, 12489 Berlin, Germany}\\ 
	\footnotesize{$^{2}$Max-Born-Institut, 12489 Berlin, Germany} 
}

\date{}

\usepackage{color}

\begin{document}

\maketitle 
\thispagestyle{firstpage}


\textbf{
	Fluktuationen sind allgegenwärtig. 
	Sie bilden einen fundamentalen Baustein der Quantenmechanik und sind verantwortlich für 
	die unterschiedlichsten Phänomene, angefangen in der Biologie bis hin zur Kosmologie.
	Fern jeder klassischen Intuition findet man zum Beispiel Kräfte zwischen 
	nichtmagnetischen und elektrisch neutralen Objekten. 
	Einzig bedingt durch Quantenfluktuationen entstehen diese Kräfte wie aus dem Nichts. 
	Seit jeher begeistern sie die Grundlagenforschung mit neuen Einsichten und 
	gewinnen immer mehr an Bedeutung für verschiedene Zukunftstechnologien.
	}\\


Mit \emph{pantha rhei} (Alles fließt), werden manchmal die Lehren des griechischen 
Philosphen Heraklit zusammengefasst. Sie entsprechen dem Verständnis von kausaler 
Verknüpfung sowie einem ständigem Wandel in der Natur.
Viele Jahrhunderte später folgte die Quantenmechanik und hat, wie wenige andere 
Theorien zuvor, unser Weltbild revolutioniert.
Die nichtintuitiven Resultate der Quantentheorie bieten 
Raum für Spekulationen und werden bis heute debattiert. 
Eines dieser Resultate ist die Erkenntnis, dass jede physikalische Größe eine 
intrinsische Unsicherheit mit sich trägt. 
Man spricht von Quantenfluktuationen, die jede Observable mit statistischem Rauschen 
verbinden. 
Heute wissen wir: \emph{Alles rauscht}.
Mehr noch, was sich wie eine theoretische Kuriosität anhört, hat weitreichende 
Implikationen im gesamten Spektrum der Physik. 
Denn die Quantenfluktuationen sorgen für messbare Effekte mit wachsender Bedeutung 
für moderne Nano- und Quantentechnologien.

Mathematisch gesehen sind Quantenfluktuationen die Konsequenz der nicht-kommutativen 
Struktur der Quantenmechanik. Vereinfacht gesagt bedeutet nicht-kommutativ für uns, 
dass die Reihenfolge von Messungen wichtig ist. 
So können wir beispielsweise niemals \glqq gleichzeitig\grqq{} die Position $x$ und 
den Impuls $p$ eines Körpers exakt bestimmen. 
Als Konsequenz ergibt sich die berühmte Unschärferelation von Heisenberg.
Im Fall einer eindimensionalen Dynamik gilt zum Beispiel für das Produkt der 
jeweiligen Unschärfen
\begin{equation*}
	\Delta x  \Delta p \ge \hbar/2,
\label{UncPrinc}
\end{equation*}
wobei $\hbar=h/(2\pi)$ die reduziete Plancksche Konstante ist. 
Je kleiner die Unschärfe im Ort $\Delta x$, desto ungenauer wird der Impuls 
$\Delta p$.

Die Unschärferelation hat einen beträchtlichen Einfluss auf unsere physikalischen 
Theorien. Zum einen muss die Statistik von Vielteilchensystemen überdacht werden: 
Sind in der klassischen Physik zwei identische Teilchen noch eindeutig durch ihre 
Position und Impuls unterscheidbar, so ist das in der Quantenmechanik durch die 
Unschärfe nicht mehr der Fall.
Zum anderen, da das Produkt der Unschärfen niemals verschwinden darf, ergibt sich, 
dass die niedrigste mittlere Energie eines Systems nicht Null sein kann.
Beispielsweise gilt für die mittlere Energie eines harmonischen Oszillators mit 
Resonanzfrequenz $\omega_0$
\begin{equation}
	\langle E \rangle
\geq \frac{\hbar\omega_0}{2}\equiv E_0.
	\label{NrgConstraint}
\end{equation}
Das entsprechende Minimum $E_0$ 
bezeichnet man als \emph{Grundzustandsenergie} bzw. \emph{Nullpunktsenergie} 
und da dieses Ergebnis unabhängig von der Temperatur ist, verschwindet das Minimum auch nicht für $T\to 0$.

Die Betrachtungen zum harmonischen Oszillator sind keineswegs Besonderheiten unseres einfachen Beispiels. 
Durch formale Ähnlichkeiten in den Bewegungsgleichungen ergeben sich Fluktuationen 
und Unschärfen auch für die Felder, die Licht und Materie beschreiben. Nehmen wir 
diesen Gedanken physikalisch ernst, so ist der Grundzustand nahezu jedes Systems 
von Unruhe gezeichnet und immer auf gewisse Art am \glqq Brodeln\grqq{}; auch 
im Vakuum und am absoluten Nullpunkt der Temperatur~\cite{Milonni94}.

Als Folge der Quantenfluktuationen
ist die Grundzustandsenergie eines Quantensystems im Allgemeinen also nie 
Null. Grundsätzlich ist diese Vorhersage nicht problematisch: Absolute Energien 
sind prinzipiell nicht messbar und echte Experimente können sich nur mit 
Energiedifferenzen auseinandersetzen. Man könnte daher vermuten, dass 
sich die Grundzustandsenergie während einer Messung immer identisch aufhebt 
und dass Quantenfluktuationen nur für Rauschen verantwortlich sind.


\section*{Fluktuationskräfte}


Allerdings hängen die Eigenschaften der Quantenfluktuationen auch von den 
Systemparametern ab. Im Fall des harmonischen Oszillators [Gl. \eqref{NrgConstraint}] 
ist es einfach die Resonanzfrequenz.
In komplexeren Systemen wird es eine Reihe von zusätzlichen Geometrie-, Material- 
und Protokollparametern sein. 
Ändert man einen der Parameter, so ändern sich die Fluktuationen und damit auch 
die Grundzustandsenergie des Systems. 
Im Experiment schlägt sich die Änderung der Grundzustandsenergie in Form 
eines von den Systemparametern abhängigen Potentials auch auf das 
absolute Messergebnis nieder.
Ist der Parameter eine Länge, muss die entsprechende Änderung als 
      fluktuations-induzierte Kraft auflösbar sein. Im Fall eines 
			Winkels ergibt sich ein Drehmoment.
			Analoge Betrachtungen gelten für weitere Paare konjugierter Größen.

In der Tat sind diese sogenannten \emph{Fluktuations-induzierten Wechselwirkungen}
heutzutage ein wichtiger und unumstrittener Bestandteil von Experimenten. 
Mehr noch, im Fall von Quantenfluktuationen sind sie unmittelbar mit der 
Unschärferelation verbunden und durch klassische Betrachtungen nicht zu erklären.

\subsection*{van-der-Waals-Kräfte}

Interessanterweise wurden die ersten Hinweise auf Quanten-Fluktuationskräfte 
schon vor der Geburt der Quantenmechanik gefunden. 
Im Jahre 1873 formulierte van der Waals seine berühmte Zustandsgleichung für 
reale Gase.
Um Phasenübergänge zu erklären, benötigte van der Waals zwei empirische 
Konstanten -- das Kovolumen und den Kohäsionsdruck. Während ersteres 
widerspiegelt, dass reale Gasatome ein gewisses endliches Volumen im Raum 
einnehmen, entspricht letzteres einer allgemeinen, anziehenden und 
nicht-kovalenten Wechselwirkung zwischen Gasmolekülen, der sogenannten 
van-der-Waals-Kraft.

Die mikroskopische Erklärung einer solchen Kraft zwischen
nichtpolaren Atomen (z.B. bei Edelgasen) wurde 1930 von F. London 
\cite{CasimirPhysics11} geliefert. 
Mithilfe einer perturbativen Rechnung 
gelang London eine Abschätzung des Kohäsionsdrucks, 
die sehr nahe an den damals bekannten experimentellen Werten lag.
London erkannte in den Berechnungen auch die wichtige Rolle der Dynamik und der dazugehörigen 
Frequenzabhängigkeiten der Systemparameter und prägte in Folge den Begriff der \emph{Dispersionskraft}
~\cite{Scheel08}. 

\begin{figure}[t!]
 \center
 \includegraphics[width=0.8\linewidth]{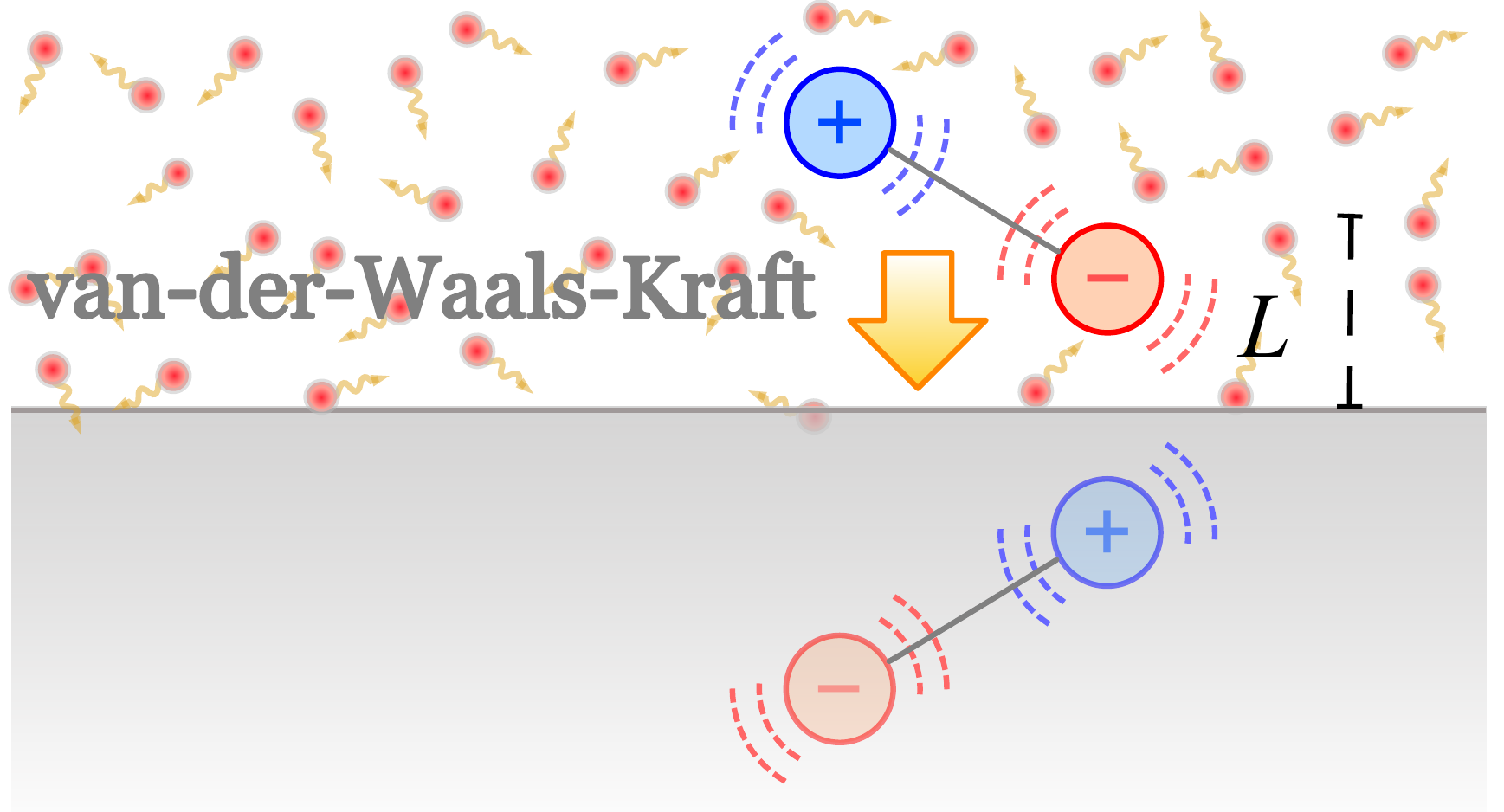}
   \caption{
	    Ein fluktuierender Dipol im Quantenvakuum vor einem Substrat wechselwirkt mit der von ihm induzierten 
	    Spiegelladung. Die Wechselwirkung erzeugt eine anziehende (van-der-Waals-)Kraft senkrecht zu der
	    Oberfläche.
			\label{fig:Dipol}
			}
\end{figure} 

Um eine direkte Verbindung zwischen der van-der-Waals-Kraft und 
den Quantenfluktuationen herzustellen ist es instruktiv, den Fall 
eines nichtpolaren Atoms im Abstand $L$ vor 
der flachen Oberfläche eines ausgedehnten und ungeladenen Halbraum zu betrachten (Abb.~\ref{fig:Dipol}). 
Obwohl die atomare Ladungswolke elektrisch neutral ist, unterliegt 
sie doch Quantenfluktuationen. 
Diese erzeugen zu einer bestimmten Zeit $t$ eine asymmetrische 
Ladungsverteilung und damit ein momentanes elektrisches Dipolmoment, 
beschrieben durch den Operator $\hat{\mathbf{d}}(t)$.
Dieser fluktuierende Dipol erzeugt einen Spiegel-Dipol im Substrat, 
dessen Feld $\hat{\mathbf{E}}$ auf das Atom rückwirkt.
Im einfachsten Fall eines perfekt leitenden Substrats erhält man 
für die elektromagnetische Energie dieses Systems 
\begin{equation}
	E = -\frac{\langle \hat{\mathbf{d}}(t)\hat{\mathbf{E}}(t)\rangle}{2}
			\approx 
			-\frac{\langle \hat{\mathbf{d}}^{2}(t)	\rangle}{48\pi \epsilon_{0}L^{3}}
	\label{London}
\end{equation}
($\epsilon_0$ Permittivität des Vakuums). Durch die  quantenmechanische Unschärfe verschwindet 
$\langle \hat{\mathbf{d}}^{2}(t)\rangle$ auch
nicht am absoluten Nullpunkt der Temperatur.
Im Gleichgewicht führt die quantenmechanische Mittelung dann zu einer 
zeitunabhängigen, aber $L$-abhängigen Energie. 
Nach Differentiation bezüglich des Abstandes ergibt sich eine anziehende 
Kraft $F\propto L^{-4}$ zwischen dem Substrat und dem Teilchen, die aus 
Symmetriegründen senkrecht zur Oberfläche wirkt.
Ähnliche Argumente führen auch zu der von London 
berechneten Abstandsabhängigkeit $F\propto L^{-7}$ der van-der-Waals-Kraft 
zwischen zwei Atomen.

Trotz ihrer -- im Vergleich zu kovalenten Bindungen -- geringen Stärke, 
spielen van-der-Waals-Kräfte eine wichtige Rolle in chemischen und 
biologischen Systemen. 
Zu den vielen Beispielen zählen Phänomene und Anwendungen im 
Bereich der Adhäsion, Reibung und Kondensation, sowie 
der komplexen Dynamik von Kolloidsystemen, aber auch die
Struktur zellularer Biomembranen~\cite{Parsegian06}.

\subsection*{Casimir und das Vakuum}

Die Untersuchung von Kolloiden stimulierte auch den folgenden theoretischen 
Fortschritt. Im Jahr 1947 entwickelten Overbeek und Verwey eine Theorie zur 
Stabilität von Kolloidallösungen, die vollständig auf London-van-der-Waals-Kräften 
basiert. 
Allerdings zeigten Experimente, dass die Stärke der Fluktuationskraft in 
Kolloidlösungen schneller mit dem Abstand fällt als von der Theorie vorhergesagt.  
Der Grund dafür, spekulierte Overbeek, könnte die vereinfachte Form des 
elektromagnetischen Feldes sein, die London in seiner Herleitung verwendet hatte. 
Tatsächlich hatte London die Endlichkeit der Lichtgeschwindigkeit ignoriert 
    und stattdessen eine instantane Ausbreitung des Feldes angenommen.

Overbeek's Vermutung inspirierte Casimir und Polder
dazu, eine relativistische Erweiterung der oben skizzierten Theorie zu formulieren. 
Am Beispiel eines Atoms vor einem Substrat fanden 
sie auf Basis einer Störungsrechnung eine Begründung für die Abweichung 
der Fluktuationskraft in Gl.~\eqref{London} für große Abstände~\cite{CasimirPhysics11}. 
Die relevante Längenskala $\lambda$ ist hier $\lambda=$ $\lambda_{\rm At}/(2\pi)$ 
mit $\lambda_{\rm At}$ der Übergangswellenlänge des charakteristischen 
atomaren Spektrums ($\lambda\sim130$ nm für Alkaliatome).
Für Abstände $L\gg\lambda$ erhält man einen Abfall der Energie 
$\propto L^{-4}$ im Vergleich zur $L^{-3}$-Skalierung in der nicht-relativistischen 
Rechnung.
Für ein perfekt leitendes Substrat ergibt sich konkreter
\begin{equation}
E\stackrel{L\gg \lambda}{\approx} 
- \frac{3\alpha_0}{32\pi^2 \epsilon_0}\frac{\hbar c}{L^4},
\label{Casimir-Polder}
\end{equation}
wobei $\alpha_0$ die statische Polarisierbarkeit des Atoms ist.
Ein ähnlicher Anstieg um eine Potenz im Abstandsgesetz findet 
sich auch im Fall der Kraft zwischen zwei 
Atomen. 
Qualitativ lässt sich der Abfall auch so erklären: Wegen der Verzögerung bei der Informationsübertragung
    durch das Feld, wird die Anti-Korrelation zwischen Dipol und Bild-Dipol 
		bei größeren Abständen zunehmend geschwächt.

Durch das vollständig quantenelektrodynamische Verfahren von Casimir und 
Polder, das erstmals auch die Grundzustandsfluktuationen des elektromagnetischen
Feldes berücksichtigte, lässt sich das Erscheinen einer abstandsabhängigen
Wechselwirking (oft Casimir-Polder-Effekt genant) auch anders interpretieren. 
Man kann eine direkte Verbindung zum Lamb-Shift herstellen, wofür
fast zeitgleich von Bethe eine entsprechende Theorie formuliert wurde. Wie in der
    ursprünglichen Casimir-Polder-Rechnung, betrachtete Bethe die Änderung der
		atomaren Übergangsenergien durch die Wechselwirkung mit dem elektromagnetischen 
		Vakuum.
Im Unterschied zu Casimir und Polder erforschte Bethe aber
ein homogenes und isotropes Vakuum ohne Oberflächen. Dennoch sind die physikalischen 
Mechanismen und technischen Probleme in beiden Fällen sehr ähnlich. 
Der von Casimir und Polder gefundene Effekt kann durch die zusätzlichen Randbedingungen 
daher auch als orts- und geometrieabhängiger Lamb-Shift verstanden werden.

Casimir selbst war wohl von dem simplen Ausdruck in Gl.~\eqref{Casimir-Polder} überrascht. 
Um dem auf den Grund zu gehen und inspiriert von einem Gespräch mit Bohr~\cite{Milonni94}, 
führte er eine konzeptionell vereinfachte, aber dafür nicht-perturbative Rechnung durch, 
die die Relevanz der Grundzustandsfluktuationen in den Vordergrund
stellt.
Er betrachtete zwei sich gegenüberstehende, neutrale und nichtmagnetische Platten (Kavität) in unterschiedlichen Konfigurationen. 
Einmal seien die Platten auf endlichem Abstand $L$ und einmal auf Abstand $L\to\infty$. 
Für das elektromagnetische Feld sind die Moden der Kavität durch die Frequenzen $\omega_{\mu}(L)$ beschrieben.
Jede der Moden verhält sich wie ein harmonischer Oszillator und trägt ihre eigene Grundzustandsenergie. 
Die gesamte minimale relative Energie berechnet sich dann via~\cite{Milonni94}
\begin{equation}
\label{Eq:SumRes}
E=
\sum_{\mu}
\frac{\hbar \omega_{\mu}(L)}{2}
-
\left[\sum_{\mu}
\frac{\hbar \omega_{\mu}(L)}{2}\right]_{L\to\infty}
\end{equation}
und entspricht effektiv der (negativen) Arbeit, die man leisten muss, um die 
Platten auf einen endlichen Abstand $L$ zusammenzuführen. 
Obwohl beide Summen in Gl.~\eqref{Eq:SumRes} divergieren, ist deren Differenz 
endlich und nicht Null~\cite{Milonni94}. 
Entgegen der naiven Vermutung, heben sich die Grundzustandsfluktuationen in 
diesem Fall nicht komplett weg.  
Die relative Grundzustandsenergie in Gl.~\eqref{Eq:SumRes} wird abstandsabhängig 
und führt zur \emph{Fluktuationskraft} $F=-\partial_L E$.
Für zwei perfekt leitende Platten der Fläche $A$, die sich parallel im Vakuum 
gegenüber stehen, sagte Casimir mithilfe von Gl.~\eqref{Eq:SumRes} die 
Energie
\begin{equation}
E=-\hbar c\frac{\pi^2}{720}\frac{A}{L^3}
\label{Casimir48}
\end{equation}
voraus. Dem entspricht eine Anziehungskraft,
die aus Symmetriegründen senkrecht zu den Oberflächen wirkt.
Abgesehen von numerischen Vorfaktoren hängt die Kraft nur von der Geometrie 
und zwei universellen Konstanten ab, die die relativistische ($c$) und die 
quantenmechanische ($\hbar$) Beschreibung des elektromagnetischen Feldes 
offenbaren. 
Die Herleitung von Casimir assoziiert die Grundzustandsfluktuationen des 
elektromagnetischen Feldes eindeutig mit einer Kraft, die heute 
seinen Namen trägt. 
Aufgrund des allgemeinen Zusammenhangs zwischen Fluktuationen und Feld-Dynamik 
findet man ähnliche Effekte auch für andere Felder, angefangen bei der 
Quantenchromodynamik bis hin zur Stringtheorie~\cite{Bordag09a}, aber auch in klassischen System, 
wie beim kritischen Casimir-Effekt ~\cite{Gambassi09}. 
Dies hat mögliche Implikationen für das Verständnis physikalischer 
      Systeme, sowohl auf mikroskopischen -- z.B. für das Innere eines Nukleons -- 
			als auch auf kosmologischen Skalen -- etwa die Genesis und Dynamik von 
			Galaxien.

\subsection*{Materie und Geometrie}

Im Gegensatz zu Gl.~\eqref{Casimir-Polder}, weist Casimirs idealisierte 
Rechnung in Gl.~\eqref{Casimir48} keine Veränderung des Potenzgesetzes 
als Funktion des Abstandes auf. 
Der Grund dafür ist, dass Casimir keine Materialeigenschaften berücksichtigt 
hat -- es fehlt also eine charakteristische Längenskala $\lambda$. 
Aber auch der Ansatz über eine Summation aller Moden in 
Form von Gl.~\eqref{Eq:SumRes} stößt schnell an seine Grenzen, 
da er Systeme mit dissipativen Materialien, für welche die Resonanz-Frequenzen 
komplex werden, nicht direkt darstellen kann.

Ein eingängiger Zugang, der Geometrie und Materialeigenschaften auf natürliche 
Weise berücksichtigt, wurde 1955 von Lifshitz gefunden. 
Lifshitz betrachtete den mittleren Strahlungsdruck auf die Platten als Funktion 
des Abstandes (Abb.~\ref{casint}) und berücksichtigte Materialeigenschaften 
über die komplexe Permittivität. 
Casimirs Resultat in Gl.~\eqref{Casimir48} folgt dann als Sonderfall in entsprechenden 
Grenzwertbetrachtungen.
Die physikalische Interpretation des Zusammenhangs zwischen Lifshitz' und Casimirs 
Rechnungen wurde von van Kampen und Schram in den Jahren 1968 und 1973 geliefert~\cite{CasimirPhysics11,Bordag09a}.

Obwohl die Berechnung über den Maxwellschen Spannungstensor von Lifshitz fast 
schon klassisch anmuten mag, so muss doch eine quantenmechanische Mittelung 
vollzogen werden.
Der \glqq Strahlungsdruck\grqq{} wird daher nicht durch klassische Felder 
hervorgerufen, sondern einzig durch den quantenstatischen Gleichgewichtszustand 
des Systems, der die Quantenfluktuationen berücksichtigt.
Das ermöglicht auch eine direkte Verbindung zwischen der 
Lifshitz-Theorie und der Thermodynamik:
Feld und Materie bilden ein offenes Quantensystem~\cite{Scheel08}, das auf 
empfindliche Weise von seiner materiellen Beschaffenheit abhängt. 
So ist es zum Beispiel möglich, die Fluktuationskräfte durch geschicktes 
Design der Geometrie und Auswahl der Materialien zu modifizieren oder 
sogar Drehmomente und Abstoßung zu realisieren 
\cite{Scheel08,Woods16,Rodriguez11,Gong21}.
Es ist sicherlich nicht überraschend, dass eine solche Vielzahl von 
Möglichkeiten einen Nährboden für die Entwicklung von semi-analytischen 
und numerischen Methoden darstellt, um immer genauere Vorhersagen für 
immer komplexere Systeme zu ermöglichen~\cite{Rodriguez11}.
Hierbei wird ein besonderer Fokus oft auf die Streueigenschaften der 
Objekte gelegt, die eine alternative Formulierung der Theorie
erlauben~\cite{CasimirPhysics11,Ingold15a}.

Die Lifshitz-Theorie erfasst alle bis jetzt beschriebenen Kräfte
      und erlaubt deren Zusammenhänge zu erläutern.
Aufgrund der gleichen physikalischen Ursache, den Quantenfluktuationen, 
beziehen sich die unterschiedlichen historischen Terminologien heutzutage 
nur noch auf die Kräfte in unterschiedlichen Situationen.
So bezeichnet man die Fluktuationskraft zwischen makroskopischen Objekten 
gewöhnlicherweise als Casimir-Kraft. 
Ersetzt man mindestens ein makroskopisches Objekt durch ein mikroskopisches, z.B ein Atom oder Molekül, dann 
ist überlicherweise von Casimir-Polder-Kräften die Rede. 
Schließlich existiert in all diesen Fällen ein Abstandsbereich bei dem 
die Retardierung vernachlässigt werden kann -- dann spricht man 
üblicherweise vom van-der-Waals-Limes.

\begin{figure} 
\center
\includegraphics[width=\linewidth]{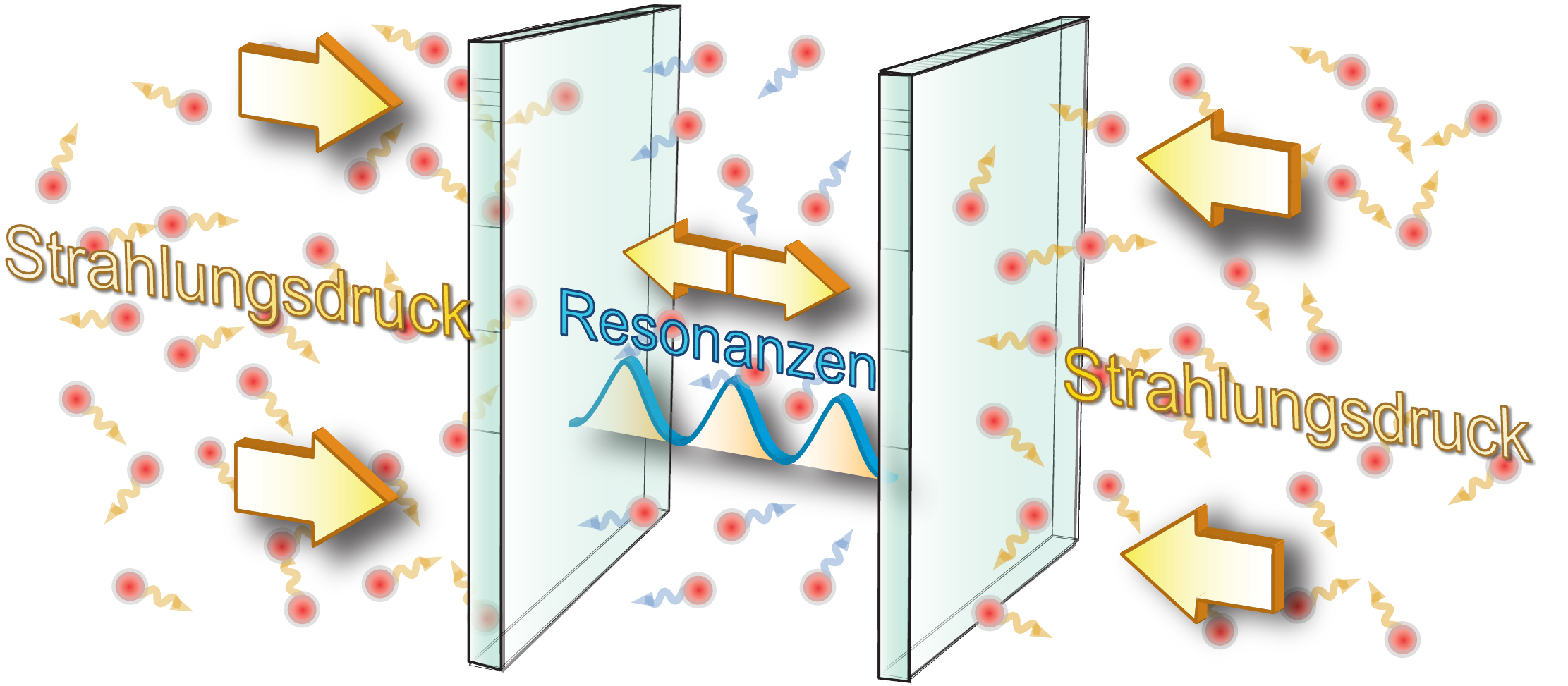}
   \caption{
			Darstellung des Casimir-Effekts über den \glqq Strahlungsdruck\grqq{} 
			der Grundzustandsfluktuationen.
			Durch die Randbedingungen an den Platten ergibt sich ein Druckunterschied, der die
			Casimir-Kraft hervorruft. 
			\label{casint}
		}
\end{figure}

\section*{Die letzten 20 Jahre im Labor}

Die experimentelle Untersuchung fluktuations-induzierter Kräfte 
      ist ein beständiger Begleiter moderner Physik. So haben Derjaguin 
			und Abrikosova schon Anfang der 1950er Jahre die van-der-Waals-Kraft 
			zwischen zwei makroskopischen Quarz-Oberflächen gemessen -- ein 
			Resultat, das nicht zuletzt auch die Entwicklung der Lifshitz-Theorie 
			stimulierte~\cite{CasimirPhysics11}.
Anfang der 1970er Jahre, folgten dann Experimente 
      an Atomen in der Nähe von Oberflächen, die entsprechende theoretische Vorhersagen 
      qualitativ bestätigten~\cite{CasimirPhysics11}.

Dank rasant voranschreitender experimenteller Möglichkeiten ist 
es in den letzten Jahrzenten einfacher 
geworden, all jene Phänomene nachzuweisen, die in der Anfangszeit der 
Casimir-Physik Ende der 1940er Jahre noch als theoretische Vorhersage 
galten.
Grund dafür sind nicht zuletzt die Fortschritte bei der Kontrolle atomarer Strahlen, der Messung von Spektraleigenschaften 
und der Laserkühlung.
So konnten bereits Anfang der 1990er Jahre die Atom-Oberflächen-Wechselwirkungen 
in verschiedensten Konfigurationen untersucht werden.
Dabei stehen z.B. Beugung, Interferenz, Übergangsfrequenzverschiebungen 
oder auch Effekte wie die Quantenreflexion im Mittelpunkt~\cite{CasimirPhysics11,Folman02}.
Ferner spielen Casimir-Polder-Kräfte in metrologischen Anwendungen eine 
wichtige Rolle, wie z.B. bei sogenannten Atom-Chips, 
wo sie sowohl für Herausforderungen also auch für Möglichkeiten einer 
      kontaktlosen Steuerung sorgen können~\cite{Folman02}.
In Atom-Chips wird eine kalte Atomwolke, beziehungsweise ein 
Bose-Einstein-Kondensat, in der Nähe von Oberflächen gefangen und für 
Präzisionsmessungen oder als Quantensensor genutzt. 
Die Casimir-Polder-Kraft begrenzt den minimalen Abstand zwischen 
      der Wolke und der Oberfläche und damit die Miniaturisierung des 
			Systems.
Zum Vergleich: Im Fall von Rubidium ist die Casimir-Polder-Kraft 
in Richtung der Oberfläche des Chips bei einem Abstand von einem 
Mikrometer schon fünf Mal stärker als die Erdanziehungskraft. 
      Bis heute sind die experimentellen Herausforderungen oft technischer Natur: 
      Eine Ansammlung von Verunreinigungen (Adsorbaten)
oder Ladungen an der Oberfläche, sowie die Oberflächenrauigkeit
beeinflussen die Stärke und Natur der Wechselwirkung.

\begin{figure} 
\center
\includegraphics[width=0.75\linewidth]{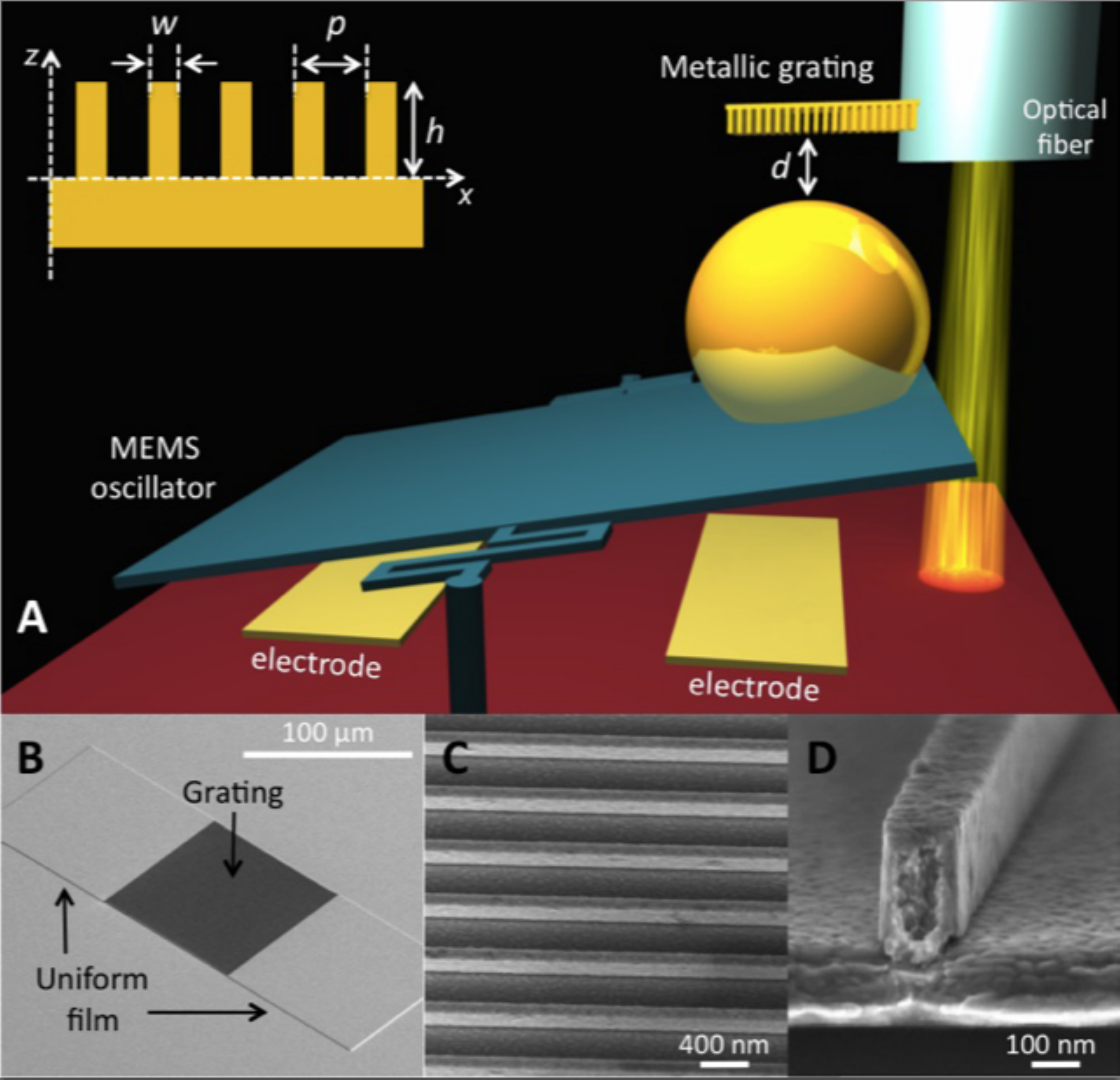}
   \caption{Darstellung eines Experiments für die Messung der Casimir-Kraft (s. Ref. \cite{Intravaia13}). Mit dem Aufbau (Bild A) bestimmt man die Verschiebung der Frequenz eines MEMS-Oszillators, die durch die Kraft zwischen einer -  am Oszillator angebrachten - Kugel und einer Gitter-Struktur (Bilder B-D) hervorgerufen wird. Für eine genaue Abstandsmessung zwischen den oszillierenden Objekten wird eine optische Faser verwendet.
			\label{casint}
		}
\end{figure}

Für eine Messung der Casimir-Kraft zwischen makroskopischen Körpern 
hingegen, mussten weitere Hürden überwunden werden.
Nicht zu unterschätzen ist die Schwierigkeit, 
zwei Platten auf Abstände von wenigen Mikrometern oder kleiner parallel 
auszurichten.
Nach ersten Messungen von Sparnaay im Jahr 1958, konnte die Casimir-Kraft 
zwischen zwei parallelen Platten erst Anfang der 2000er Jahre mit einer 
Genauigkeit von 15\% bestätigt werden~\cite{Onofrio06}.
Um das Problem der Parallelität im Experiment zu umgehen, behilft man 
sich oft anderer Geometrien, wie zum Beispiel zweier Kugeln oder einer 
Kugel vor einer Oberfläche.
Als Beispiele sind hier Experimente zum Test des Gravitationsgesetzes 
\cite{Onofrio06} sowie Nano- oder Mikro-Elektro-Mechanische-Systeme 
(NEMS und MEMS) zu nennen~\cite{Gong21}, die man heute sogar als 
Beschleunigungssensoren in Smartphones findet.
Bei diesen Systemen bewegen sich Bauteile in sehr geringem Abstand 
voneinander, wodurch Fluktuationskräfte an Bedeutung gewinnen. 
Wie auch in Atom-Chips, begrenzen die Fluktuationskräfte aufgrund
    ihrer Stärke sowohl eine genaue Messung der Schwerkraft als auch
		die Miniaturisierung und damit energetisch und sensorisch effizientere Designs.
Zum Vergleich: Wenn $A\sim 1 \mathrm{cm}^2$ und $L\sim 1 \mu$m, dann 
ist die Kraft in Gl.~\eqref{Casimir48} $0.1 \mu$N. Das entspricht 
ungefähr der Gewichtskraft von 10 $\mu$g an der Erdoberfläche.
So konnte ein erster experimenteller Nachweis der Casimir-Kraft in 
gekrümmten Geometrien schon 1996 von  Lamoreaux mit einer Gravitationswaage 
erreicht und später mithilfe eines Rasterkraftmikroskops oder MEMS mit 
höherer Genauigkeit bestätigt werden~\cite{CasimirPhysics11,Bordag09a}.
Das birgt aber die Schwierigkeit, die Krümmung der Geometrie in der 
Modellierung korrekt darzustellen.
Besondere Wichtigkeit erlangte dabei die von Derjaguin entwickelte 
\emph{proximity force approximation}, die es erlaubt eine gekrümmte 
Oberfläche durch planare Ergebnisse näherungsweise zu erfassen 
\cite{CasimirPhysics11,Bordag09a,Rodriguez11, Kardar99,Intravaia13}.

In Kombination mit den jeweiligen Materialeigenschaften führen 
Präzisionsmessungen der Casimir-Kraft 
bis heute zu ungeklärten Kontroversen.
Ein prominentes Beispiel ist die sogenannte Plasma-Drude-Kontroverse, die 
sich mit der Rolle der elektrischen Leitfähigkeit in der Lifshitz-Theorie 
befasst~\cite{Bordag09a}.
So haben Messungen bei endlicher Temperatur zwischen leitenden Oberflächen 
signifikante Abweichungen von den theoretischen Vorhersagen ergeben, wenn 
in der Modellierung die Metalle mit einer endlichen Dissipation, z.B. 
im Rahmen des Drude-Modells, beschrieben wurden. 
Das Überraschende ist, dass diese Abweichung in den meisten der betroffenen 
Experimenten verschwindet, wenn man im Drude-Modell die Dissipation künstlich 
auf Null setzt (sogenanntes Plasma-Modell eines Metalls).
Die Kontroverse entsteht daraus, dass man eigentlich das Drude-Modell 
dem Plasma-Modell vorziehen würde, da ersteres mit dem Ohm'schen Gesetz 
vereinbar ist.
Trotz einer Reihe von theoretischen Vorschlägen über die physikalischen 
Ursachen und wiederholten Experimenten, gibt es bis heute keinen Konsenz 
über die Lösung des Rätsels.


\section*{Jenseits des Gleichgewichts}

Eine weitreichende Theorie der Fluktuationskräfte wurde 1961 
von Lifshitz zusammen mit Dzyaloshinskii und Pitaevskii formuliert, 
wobei das Zusammenspiel von statistischer Physik, Quantenelektrodynamik 
und Festkörperphysik im Vordergrund steht ~\cite{CasimirPhysics11}.
Wie in den bisherigen Betrachtungen ist dabei die zentrale Annahme 
die Existenz eines thermischen Gleichgewichts.
Das vereinfacht die Rechnungen bedeutend, 
da die Korrelationsfunktionen durch die Antwortfunktionen des Systems 
exakt bestimmt sind.
Letzteres wird gemeinhin über das sogenannte Fluktuations-Dissipations-Theorem 
formalisiert.
Im Nichtgleichgewicht, hingegen, ist die Beschreibung meist deutlich 
schwieriger. In vollster Allgemeinheit ist eine derartige Beschreibung 
vermutlich sogar unmöglich, was die Entwicklung einer Zahl von Näherungsmethoden 
befördert hat.

Zu den verbreitetsten Formen von Nichtgleichgewicht bei 
fluktuations-induzierten Phänomenen zählen vermutlich Temperaturunterschiede 
zwischen räumlich getrennten Teilen des Systems~\cite{Volokitin07}. 
Dieses sogenannte \emph{thermische Nichtgleichgewicht} wird oftmals 
in zeitlich stationären Situationen betrachtet, sodass die vorliegenden 
Temperaturgradienten zu einem konstanten Energiestrom und den dazugehörigen 
Kräften im System führen. 
Insbesondere wenn die makroskopischen Objekte durch das Vakuum getrennt 
sind, werden diese Effekte ausschließlich durch die Fluktuationen des 
elektromagnetischen Feldes vermittelt. 
Interessante Nichtgleichgewichtseffekte, insbesondere im Bereich 
      des Wärmetransports aber auch der Fluktuationskräfte, können 
			vorhergesagt und experimentell untersucht werden. Deren Stärke und 
			Vorzeichen, auch in Kombination mit Materialeigenschaften, können 
			sich durch den Temperaturunterschied steuern lassen \cite{Biehs21}.

Im thermischen Nichtgleichgewicht spielen Quantenfluktuationen meinst eine 
untergeordnete Rolle und thermische Fluktuationen bestimmen weitgehend 
die Stärke des Vorgangs. 
Dies ändert sich, wenn sich unterschiedliche Teile des Systems in Bewegung 
befinden. In diesen \emph{mechanischen Nichtgleichgewichten} unterscheidet 
man grob nach den verschiedenen Möglichkeiten die Lorentzinvarianz zu brechen.
Hat man etwa ein oszillierendes ungeladenes und nichtmagnetisches Objekt 
im Vakuum, z.B. eine Platte oder ein Atom, so spricht man in der Regel 
vom dynamischen Casimir-Effekt. Dabei wird eine verschränkte Strahlung 
-- zusammen mit einer entsprechenden Rückstroßkraft -- erzeugt, 
die als eine Art parametrische Verstärkung der Grundzustandsfluktuationen 
verstanden werden kann~\cite{Kardar99,Nation12}.
Betrachtet man hingegen die Bewegung durch das elektromagnetische Vakuum 
mit \emph{konstanter} Beschleunigung, kommt es zum Fulling-Davies-DeWitt-Unruh 
Effekt, wobei das Objekt im mitbewegten Bezugssystem anstatt des Vakuums 
ein thermisches Feld mit konstanter Temperatur wahrnimmt -- ein Effekt der mit der Hawking-Strahlung eines schwarzen Lochs verwandt
ist~\cite{Nation12}. 
Die Untersuchung dieser Phänomene ist nicht nur als Test der 
      Fundamente der Quanten- und Gravitationstheorie relevant, sondern 
			trägt auch zu einem tieferen Verständnis der Dynamik ultra-kalter Atome 
			oder superleitender Schaltkreise bei, wo analoge Effekte bereits 
			gemessen worden sind~\cite{Nation12}.

%
\begin{figure}[t!]
 \center
 \includegraphics[width=0.75\linewidth]{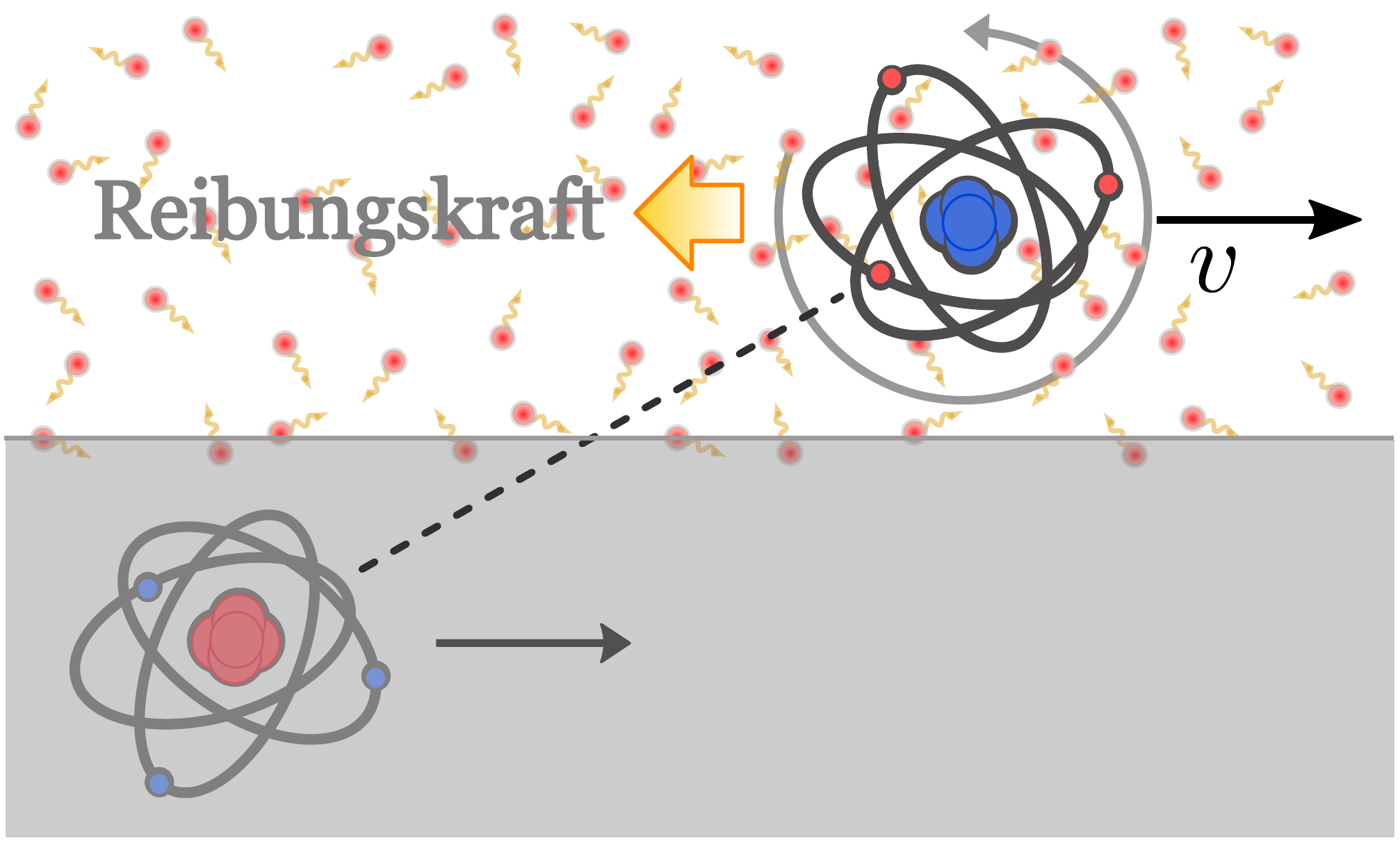}
   \caption{
	    Ein Atom im Quantenvakuum bewegt sich mit konstanter Geschwindigkeit parallel zu einem Substrat und 
	    wechselwirkt mit der von ihm induzierten Spiegelladung. Die Wechselwirkung erzeugt eine Reibungskraft parallel
	    zu der Oberfläche und eine ungewöhnliche Übertragung von Drehmoment zum Atom.
			\label{fig:MDipol}
			}
\end{figure} 
%

Ein interessanter dynamischer Effekt tritt aber auch dann auf, wenn sich 
zwei (oder mehrere) eng benachbarte Objekte mit 
\emph{konstanter Relativgeschwindigkeit} bewegen.
Im Gegensatz zur Bewegung im absoluten Vakuum liegt hierbei 
der Fokus auf der Geometrie und dem Material des Systems. Bemüht man für 
ein sich bewegendes Atom erneut das qualitative Bild des Spiegel-Dipols 
(Abb.~\ref{fig:Dipol}), so ist die Bewegung des Bild-Dipols im Vergleich 
zu einer parallelen Bewegung des Teilchens relativ zum Substrat verzögert (Abb.~\ref{fig:MDipol}). 
Der Grund dafür sind Dissipation und Dispersion innerhalb des Körpers.
Damit erhält die anziehende Kraft zwischen Dipol und Bild einen 
nichtkonservativen Anteil parallel zur Oberfläche, welcher der 
Bewegung entgegen wirkt. Ohne Quantenfluktuation würde diese Kraft 
nicht existieren: Man spricht dann von Quantenreibung~\cite{Volokitin07}.

Die Physik der Quantenreibung ist sehr vielfältig. Sie zeigt wichtige 
Verbindungen sowohl zu besonderen Symmetrie- und Materialeigenschaften 
des Systems~\cite{reiche21f}, als auch
zu den obengenannten Nichtgleichgewichtseffekten und zum Cherenkov-Effekt.
Das führt zu interessanten Phänomenen, wie zum Beispiel einer ungewöhnlichen Übertragung von Drehmoment zum Atom~\cite{reiche21f} (Abb.~\ref{fig:MDipol}).
Bemerkenswert ist, dass lokale Gleichgewichtsmethoden bei der Beschreibung, 
wie die Quantenreibungskraft auf Atome wirkt, scheitern~\cite{reiche21f}.
Um dies zu motivieren, betrachten wir die stationäre Bewegung eines Atoms 
entlang einer planaren Oberfläche mit konstantem Abstand $L$ und 
Geschwindigkeit $v$. 
Bei $T=0$ ergibt sich die Reibungskraft
\begin{align}
F
&\approx-\frac{36}{\pi^{3}}\hbar\alpha^{2}_{0}\rho^{2}\frac{v^{3}}{(2L)^{10}},
\label{rollingFric}
\end{align}
wobei $\rho$ mit der Zustandsdichte des elektromagnetischen Feldes bei 
kleinen Frequenzen verbunden ist~\cite{reiche21f}. 
Für leitende Substrate ist $\rho$ effektiv der spezifische Widerstand 
des Materials. 
Im Gegensatz zu Dispersionskräften im Gleichgewicht, die sich aus einem 
breiten Frequenzspektrum herleiten,  sind für Quantenreibung besonders 
die niedrigen Frequenzen fern jeder Resonanz im System relevant. 
Das legt physikalisch einen starken Fokus auf die dissipativen Eigenschaften 
des Systems. 
Mathematisch bedeutet dies, dass insbesondere jene Korrelationen zwischen System und Umgebung für die Beschreibung relevant sind, die von vielen geläufigen Nährungsmethoden nicht mehr erfasst werden können. Die Konsequenzen können weitreichend sein, bis hin einer Verletzung der Energiebilanz im System~\cite{reiche21f}.
Daher muss man einige Vorsicht walten lassen, um die tiefen Frequenzen adäquat zu berücksichtigen~\cite{reiche21f}.

Gleichung \eqref{rollingFric} liefert eine gute Beschreibung solange 
$v/L$ ($< 150$ $\mu$eV für typische Werte) kleiner als jede Resonanz 
im System ist. In allen anderen Fällen kann sich Gl.~\eqref{rollingFric} 
ändern~\cite{Volokitin07}. 
Insgesamt ergeben sich meistens vergleichsweise kleine Werte für 
die Kraft der Quantenreibung, die sich auch deshalb bis heute einem 
direkten experimentellen Nachweis entzogen hat.
Dafür werden aktuell verschiedene experimentelle Protokolle diskutiert.
Eine Idee wäre es, der notwendigen Genauigkeit der Messung mit der extremen Sensitivität von modernen Atom-Interferometern zu begegnen. 
Alternativ könnte man hohe Geschwindigkeiten und kleine Abstände  
durch Beugung von Atomstrahlen an feinen Gittern erreichen. 
Oder man arbeitet mit möglicherweise analogen System, die z.B. das Atom durch ein sogenanntes \emph{nitrogen-vacancy center} an der Spitze eines Rastermikroskopes~\cite{reiche21f} bzw. die Relativbewegung durch einen elektrischen Strom emulieren~\cite{Volokitin07,reiche21f}.
Solche Vorschläge haben ihre individuellen Vor- und Nachteile~\cite{reiche21f}, sodass ein direkter Nachweis nach wie vor eine Herausforderung darstellt.

Allerdings kann sich das durch geschickte Optimierungen der Systemarchitektur 
in Zukunft ändern. 
So sind Quantenreibungskräfte auf Atome hochgradig nicht-additiv. 
Die Nicht-Additivität lässt sich bereits anhand der quadratischen 
Abhängigkeit von der elektromagnetischen Zustandsdichte, d.h. 
$F\propto\rho^2$, erahnen. 
Beachten wir die enge Verbindung zwischen der Zustandsdichte und der 
Materialverteilung im System, so liegt hier ein Optimierungspotenzial, 
das es zu heben gilt.
Mit anderen Worten: Die doppelte Anzahl an wechselwirkenden Flächen 
bedeutet keineswegs die doppelte Kraft, sondern es können sich durchaus 
Überhöhungen von einigen Größenordnungen ergeben~\cite{reiche21f}.\\


\section*{Ausblick}


Fluktuationskräfte zeigen, dass sich Quantenfluktuationen im 
System nicht tatenlos verhalten, sondern auf subtile und faszinierende 
Art in seine Dynamik eingreifen. 
Schwinger nannte die Casimir-Kraft \glqq eine der am wenigsten intuitiven 
Konsequenzen der Quantenelektrodynamik\grqq{}. 
Ob intuitiv oder nicht, die Physik der fluktuations-induzierten 
Phänomene -- aus historischen Gründen oft auch  \emph{Casimir-Physik} 
genannt -- ist spätestens heutzutage ein fester Bestandteil unseres 
physikalischen Verständnisses. Sie ist von großer und 
wachsender Bedeutung für viele Untersuchungen und Anwendungen 
in den verschiedensten Forschungsfeldern, angefangen bei Nano- und 
optischen Quantentechnologien, über die Biophysik, bis hin 
zur Dynamik des Universums. 
Besonders bei den Vorhersagen im Nichtgleichgewicht stehen wir sowohl 
aus experimenteller als auch aus theoretischer Sicht noch am Anfang.

 \thispagestyle{empty}

\begin{small}

\end{small}

\newpage
\thispagestyle{empty}
\begin{figure*}[t!]
\textbf{\Large Die Autoren}\\
\HorRule\\
\\
\includegraphics[width=3cm]{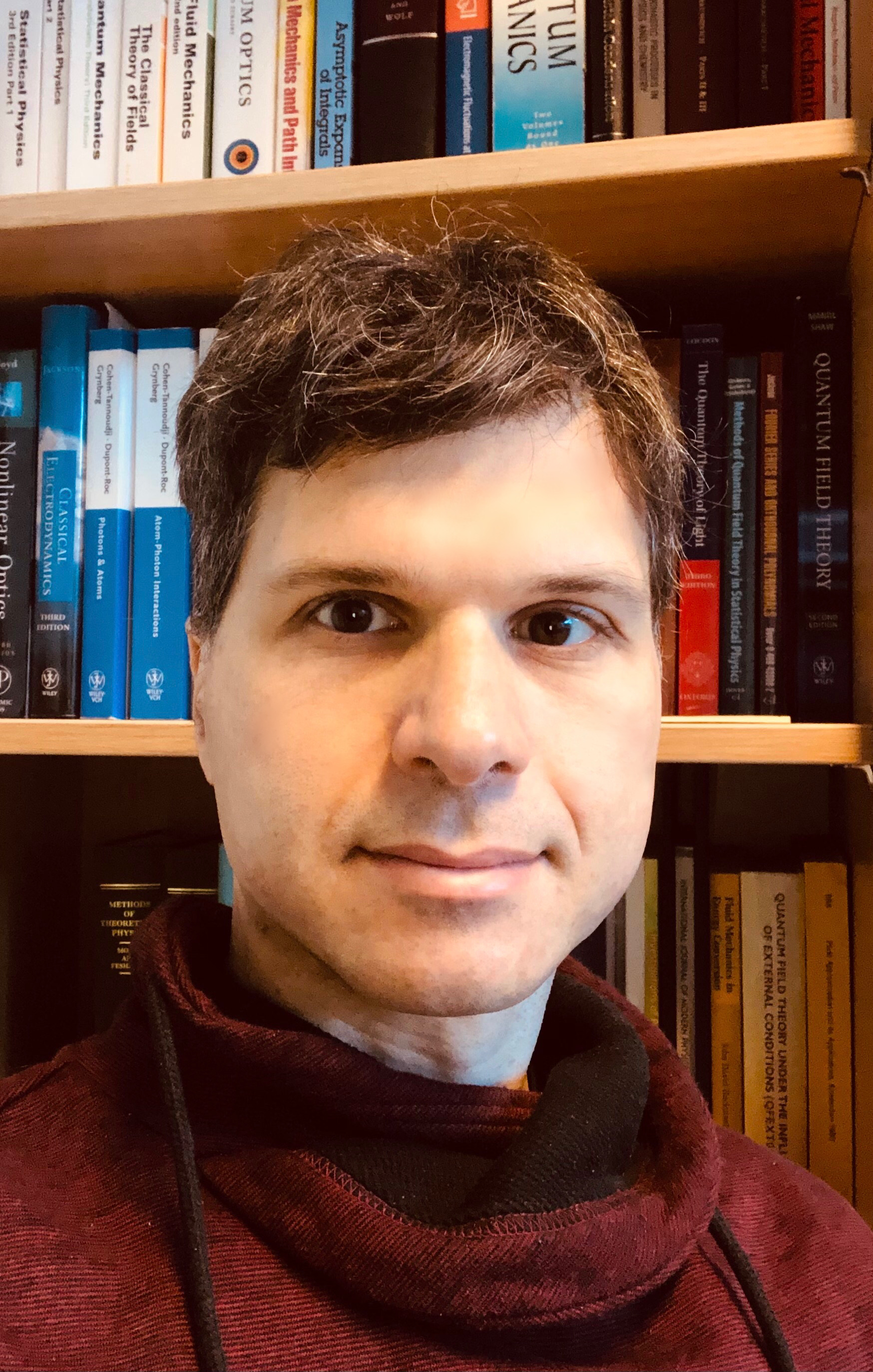}
\parbox{13cm}{\vspace{-4.5cm} \textbf{Francesco Intravaia} studierte Physik an der Università degli Studi di Palermo in Italien und promovierte in Paris an der Université Pierre et Marie Curie mit einer Arbeit über den Casimir-Effekt am Laboratoire Kastler-Brossel. Anschließend forschte er im Rahmen eines Alexander-von-Humboldt Stipendiums an der Universität Potsdam sowie als Director’s Postdoctoral Fellow am Los Alamos National Laboratory. Seit 2019 ist er akademischer Rat an der Humboldt Universität zu Berlin, wo er weiterhin Fluktuations-Induzierte Phänomene untersucht.}
\includegraphics[width=3cm]{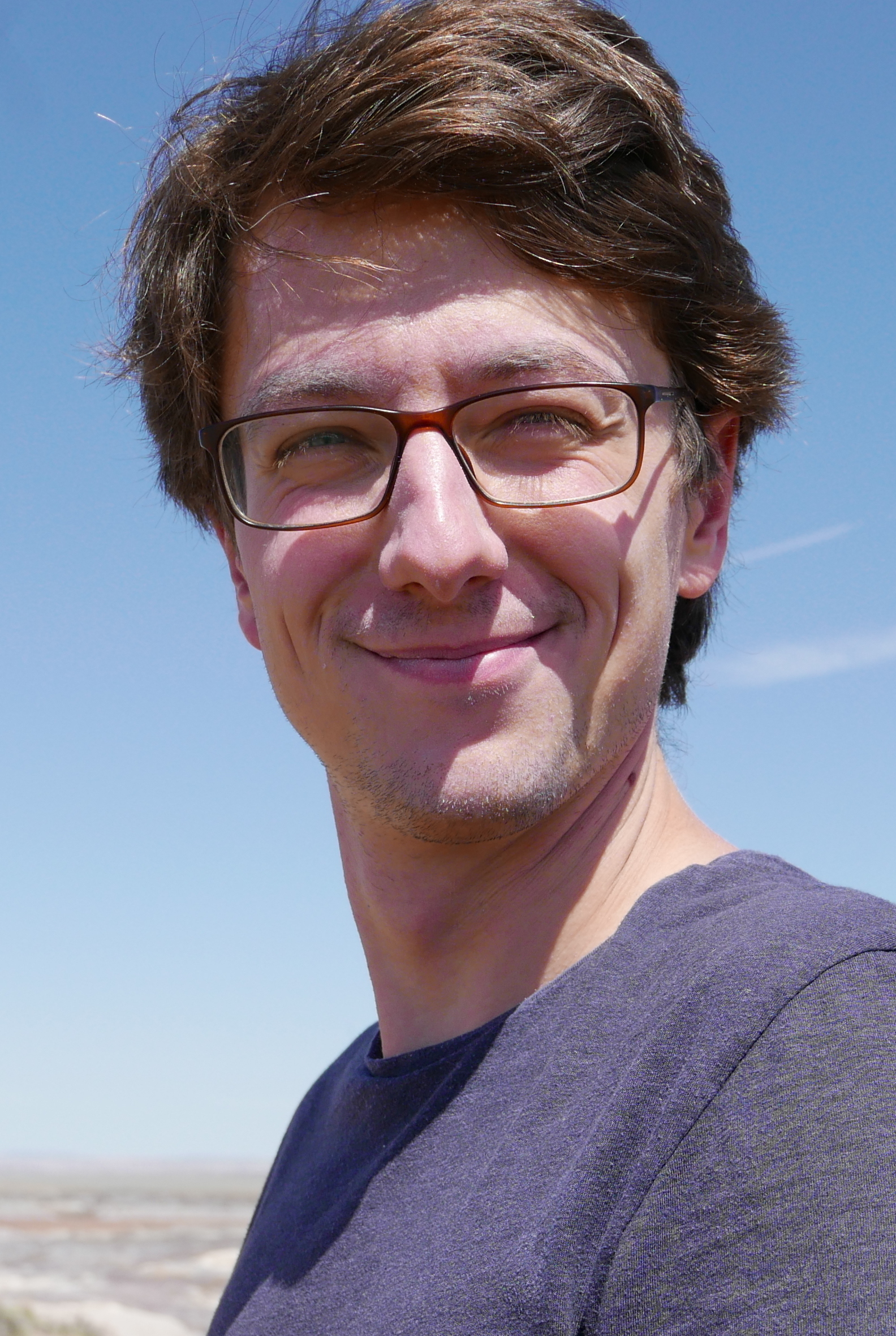}
\parbox{13cm}{\vspace{-4.5cm} \textbf{Daniel Reiche} studierte Physik in Jena und Berlin und forschte für seine Masterarbeit am Los Alamos National Laboratory. Anschließend arbeitete er am Max-Born-Institut sowie als Fulbright Stipendiat an der University of Maryland, College Park, und der Northern Arizona University. 2021 promovierte er an der Humboldt-Universität zu Berlin und arbeitet an selbiger derzeit als Post-Doc zur Miniaturisierung von Quantensensoren.}
\includegraphics[width=3cm]{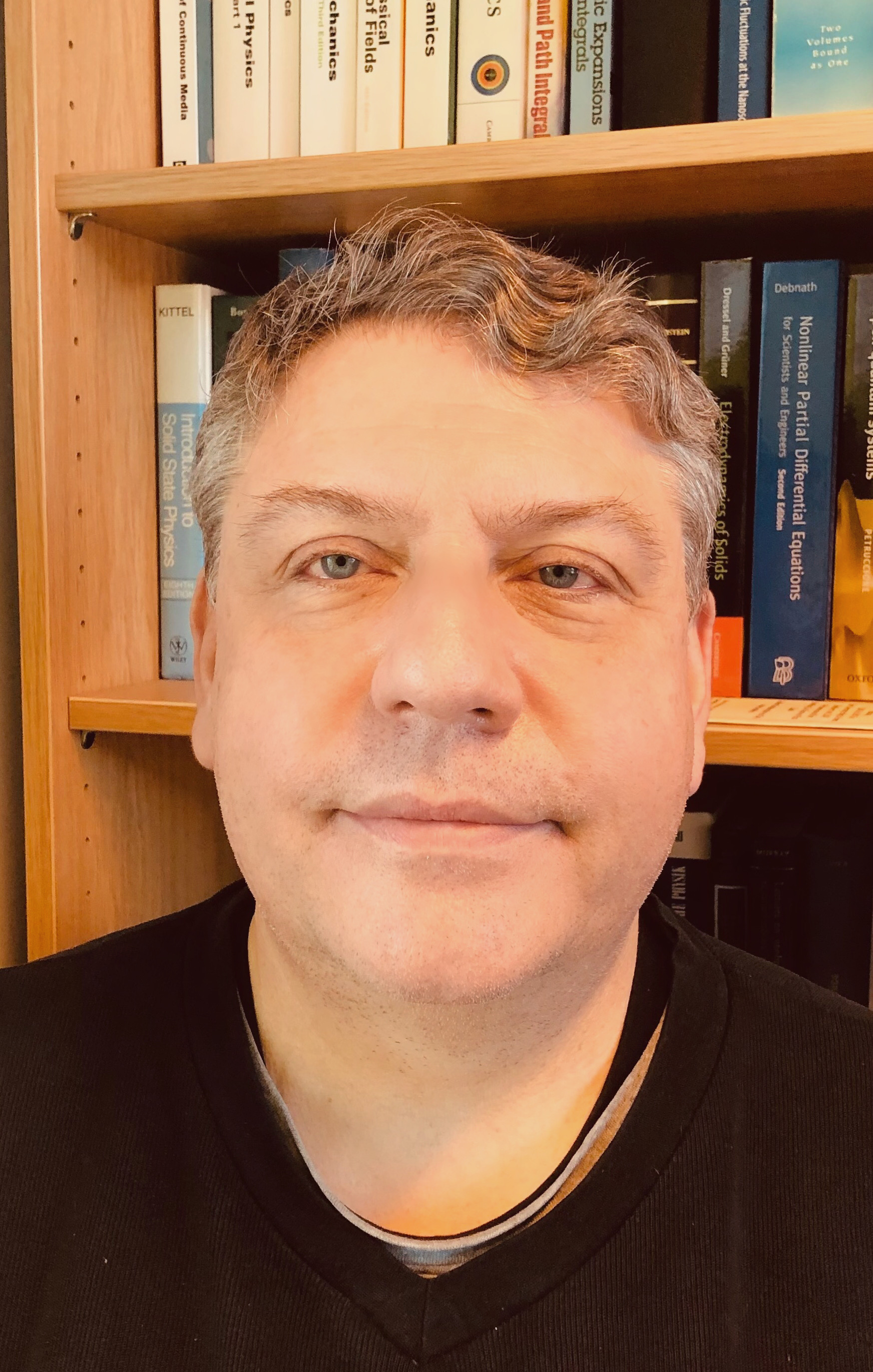}
\parbox{13cm}{\vspace{-4.5cm} \textbf{Kurt Busch} studierte Physik an der Universität Karlsruhe (TH) und promovierte an der Universität Karlsruhe (TH) sowie der Iowa State University zur Ausbreitung klassischer Wellen in ungeordneten Systemen. Nach einem PostDoc-Aufenthalt an der University of Toronto war er an der Universität Karlsruhe (TH) Nachwuchsgruppenleiter im Rahmen des Emmy-Noether Programms der DFG. Als Associate Professor forschte und lehrte er an der University of Central Florida bevor er an das Karlsruher Institut für Technologie berufen wurde. Seit 2011 ist er Professor für Theoretische Optik \& Photonik an der Humboldt Universität zu Berlin und gleichzeitig leitet er eine Arbeitsgruppe am Max-Born-Institut in Berlin.}
\HorRule
\end{figure*}

\end{document}